\documentclass[reprint, aps,superscriptaddress]{revtex4-2}
\usepackage{graphicx}
\usepackage{dcolumn}
\usepackage{bm}

\usepackage{epsfig,latexsym,cancel,amssymb,amsmath,bm}
\usepackage{graphicx}
\usepackage{feynmp}
\usepackage{subfigure}
\usepackage{epstopdf}
\usepackage{color}
\usepackage{mathtools}
\usepackage{bigints}
\usepackage[utf8]{inputenc}
\usepackage[english]{babel}
\usepackage[dvipsnames]{xcolor}
\usepackage{microtype}

\graphicspath{{Images/}}

\def\f{\phi}

\newcommand{\be}{\begin{eqnarray}}
\newcommand{\ee}{\end{eqnarray}}

\def\to{\rightarrow}

\begin{document}
 
\preprint{APS/123-QED}

\title{How to grow a flat leaf}	
\author{Salem al-Mosleh}
\affiliation{John A. Paulson School of Engineering and Applied Sciences, Harvard University, Cambridge, MA 02138, USA}
\author{L. Mahadevan}
\email{lmahadev@g.harvard.edu}
\affiliation{John A. Paulson School of Engineering and Applied Sciences, Harvard University, Cambridge, MA 02138, USA}
\affiliation{Departments of Physics, and Organismic and Evolutionary Biology, Harvard University, Cambridge, MA, USA}

\date{\today}

\begin{abstract} 
Growing a flat lamina such as a leaf is almost impossible without some feedback to stabilize long wavelength modes that are easy to trigger since they are energetically cheap. Here we combine the physics of thin elastic plates with feedback control theory to explore how a leaf can remain flat while growing. We investigate both in-plane (metric) and out-of-plane (curvature) growth variation and account for both local and nonlocal feedback laws.  We show that  a linearized feedback theory that accounts for both spatially nonlocal and temporally delayed effects suffices to suppress long wavelength fluctuations effectively and explains recently observed statistical features of growth in tobacco leaves. Our work provides a framework for understanding the regulation of the shape of leaves and other laminar objects. 
\end{abstract}
\maketitle

Shape is an emergent property of matter enabling function at every level in biology, from the molecular to the organismal. To ensure the robust generation of shape \cite{plant-stochasticity} in the (unavoidable) presence of noise, feedback mechanisms must couple sensing and growth \cite{robustness-plants,plant-read-shape, 2021roadmap,flucts-pcepshn-sci}. 
This can be seen in the garden in plant leaves that are often flat\cite{conformal-leaf1, conformal-leaf2}, a configuration that is hard to achieve in thin growing laminae without feedback as they are susceptible to bending. In fact, recent studies in \textit{N. tabacum} (tobacco) leaves show spatially correlated fluctuations of areal growth rate \cite{sharon-2020}, consistent with an important role for feedback in maintaining this shape \cite{hamant2008science, meyerowitz2014, hamant2019microtubules, coen-2020-microtubles, stochastic-growth-lubensky}.

\begin{figure} [!t]
    \centering
    \includegraphics[width=0.9\columnwidth]{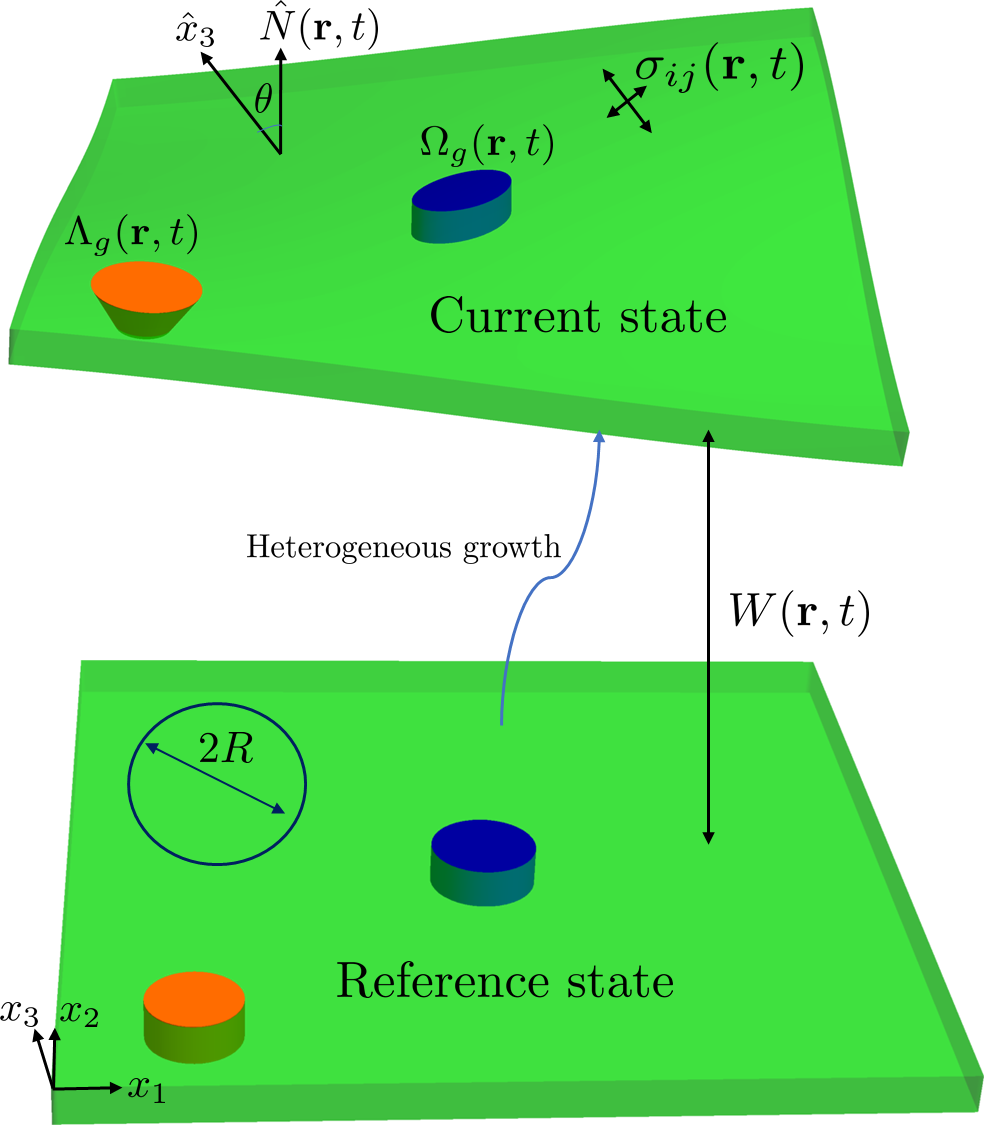}
    \caption{\textbf{Schematic of a growing, nearly flat, thin elastic plate.} The shape of the leaf results from the inhomogeneous growth both in the in-plane (encoded by $\Omega_g(\mathbf{r}, t)$) and out-of-plane (encoded by $\Lambda_g(\mathbf{r}, t)$) directions. The  shell is described by the out of plane deflection $W(\mathbf{r}, t)$ and the scaled Airy stress function (see text). The unit vector $\hat{x}_3$ points in the vertical direction and $\hat{N}(\mathbf{r}, t)$ is the normal to the surface. The angle between them is $\theta(\mathbf{r}, t)$. $R$ is the size of a region over which we coarse-grain measurement of growth.}
    \label{fig:problem-setup}
\end{figure}

Motivated by these observations, here we propose a framework within which to study the control of thin surfaces, modeled as thin elastic plates that grow and change over time. We focus on small deviations from the flat state and study in-plane and across thickness growth (Fig.~\ref{fig:problem-setup}) and show that strain \cite{shraiman2005mechanical} and curvature \cite{pulwicki2016plants, growth-of-form} sensing together stabilize the flat state. 


\textit{Mechanics of a growing lamina.}  We assume that the elastic plate (see SI for a generalization to growing shells with growth rate g), has constant thickness $h$  and a mid-surface parameterized using Cartesian coordinates $\mathbf{r} = (x_1, x_2)$ and time $t$, and the a deflection $W(\mathbf{r}, t)$ of the mid-surface above the reference $(x_1, x_2)$-plane (Fig.~\ref{fig:problem-setup}). 
For an isotropic material, there are two elastic constants, $E,\nu$ in terms of which we can write a stretching stiffness $S=Eh/12(1-\nu^2)$ and a bending stiffness $B=Eh^3/12(1-\nu^2)$ that link the strain measures to the stress tensor $\sigma_{ij}(\mathbf{r}, t)$.
The scaled Airy stress function, $\phi(\mathbf{r}, t)$, is related to the stress tensor through $\sigma_{11} = E \partial_{2}\partial_{2} \f, \;\sigma_{22} = E \partial_{1}\partial_{1} \f$ and $\sigma_{12} = -E \partial_{1}\partial_{2} \f$. The Laplacian of the scaled Airy stress function satisfies $E \triangle \phi(\mathbf{r}, t) = \sigma_{11} + \sigma_{22}$ and is proportional to the areal strain. 

Since growth-induced elastic frustration leads to rapid equilibration (at the speed of sound) while feedback from strains that modulates growth is generally slow (due to the time scales over which these signals are transduced biochemically), the in-plane and out-of-plane growth are treated quasi-statically. Then, at linear order we can describe an elastic plate using the depth-averaged compatibility and transverse force balance equations \cite{long-leaf} as 
\begin{eqnarray}
	\triangle^2 \f(\mathbf{r}, t) & = & \Omega_g(\mathbf{r}, t) 
	\label{eq:plate-eqs1} \\
	h\;\triangle^2 W(\mathbf{r}, t) & = & \Lambda_g(\mathbf{r}, t),
	 \label{eq:plate-eqs2}
\end{eqnarray}
where $\Omega_g(\mathbf{r}, t)$ reflects the incompatibility due to in-plane growth, and $\Lambda_g(\mathbf{r}, t)$ is the pressure induced by variations in growth through the thickness of the plate (see Fig.~\ref{fig:problem-setup} and \cite{long-leaf}).  
%

\textit{Local and instantaneous feedback.} {Our goal is to study the stability of the flat, stress-free state.} At linear order, the most general form of the feedback law that we can write to account for the slow feedback from shape to growth leads to   (see SI for details of other equivalent forms written in terms of geometric quantities such as the first and second fundamental forms)
\begin{eqnarray}
	 &\partial_t \Omega_g = \int_{-\infty}^{t} dt^\prime d^2\mathbf{r}^{\prime} \left[G_{11} \triangle^2 \phi +  G_{12} \;h \triangle^2 W(\mathbf{r}^{\prime}, t^{\prime}) \right], & \;\;\;\;\;\; \label{eq:kernel-growth-1}
	\\
	&\partial_t \Lambda_g =  \int_{-\infty}^{t} dt^\prime d^2\mathbf{r}^{\prime} \left[G_{21} \triangle^2 \phi +  G_{22} \;h \triangle^2 W(\mathbf{r}^{\prime}, t^{\prime}) \right]. &\;\;\;\;\;\; \label{eq:kernel-growth-2}
\end{eqnarray} 
where $G_{ij}(\mathbf{r}  - \mathbf{r}^{\prime}, t - t^{\prime})$ are the feedback kernels that represents the (possibly non-local and delayed) coupling of growth to the state of the system described in terms of the inhomogeneous in-plane growth $\triangle^2 \f(\mathbf{r}^{\prime}, t^{\prime})$ (analogous to the Gauss curvature or the Ricci scalar) and $\triangle^2 W(\mathbf{r}^{\prime}, t^{\prime})$ (analogous to a transverse pressure). We note that dimensional analysis demands that the coefficient of $\triangle^2 W$ be a factor $O(h)$ smaller than the growth rate of the corresponding perturbation of $\triangle^2 \phi$.

The simplest choice of kernel corresponds to the case of local feedback --- where the instantaneous rate of expansion and shear due to growth is a function of local variables such as curvature and strain (see SI)--- and is described using the Dirac delta function by 
\begin{eqnarray}
	G_{ij}(\boldsymbol{\rho}, \tau) = -\alpha_{ij} \;\delta(\tau)\; \delta(\boldsymbol{\rho}), \label{eq:kernel-local}
\end{eqnarray}
where the feedback matrix $\alpha_{ij}$, with units of time$^{-1}$, is not required to be symmetric. 
To understand the stability of a nominally flat plate to perturbations of growth, we substitute the ansatz $\phi(\mathbf{r}, t) = \phi_0 e^{i \textbf{q}\cdot \textbf{r} + \lambda t}$ and $W(\mathbf{r}, t) = W_0 e^{i \textbf{q}\cdot \textbf{r} + \lambda t}$ into Eqs.~(\ref{eq:plate-eqs1}-\ref{eq:kernel-local}), which gives two eigenvalue solutions for the growth, given by
\begin{eqnarray}
	  \lambda_{\pm}(\mathbf{q}) = -\frac{1}{2}\left(tr[\alpha] \pm \sqrt{tr[\alpha]^2 - 4 \;det[\alpha]}\right),
	\label{eq:local-rates}
\end{eqnarray}
where $tr[\alpha]$ and $det[\alpha]$ are the trace and determinant of the feedback matrix $\alpha_{ij}$. Note that the growth rates are independent of wavenumber $\mathbf{q}$, and therefore independent of system size $L$, consistent with the local nature of the growth law. 

The system will be stable when $ Re \lambda_{\pm}(\mathbf{q}) <0$, which happens when $det[\alpha] > 0, tr[\alpha] > 0$.  When $4 det[\alpha] > tr[\alpha]^2$ the system will be oscillatory but still stable. We note that stability requires both curvature and strain sensing, which ensures that $\det[\alpha] \neq 0$, and the lamina needs to be able to measure/sense deviations from the target flat, stress-free state driven by both curvature and in-plane growth.

However, while local and instantaneous feedback can lead to asymptotic stability of the flat state, we now show that this form of feedback is not efficient at suppressing stochastic fluctuations at the scale of the system size $L$. To see this, we extend Eqs.~(\ref{eq:kernel-growth-1} - \ref{eq:kernel-growth-2}) and add stochastic terms to represent the effects of fluctuations in in-plane and curvature growth rate. We assume that fluctuations in in-plane areal growth, written as $\triangle \chi_\f(\mathbf{r}, t)$, and curvature growth, written as $\triangle \chi_W(\mathbf{r}, t)$ that appear through the variations in the metric and curvature tensor (see SI) have a white noise spectrum   so that
\begin{eqnarray}
    &~&\langle \triangle \chi_\f(\mathbf{r}, t)  \triangle \chi_\f(\mathbf{r}^\prime, t^\prime)\rangle = D_\f \delta(\mathbf{r} - \mathbf{r}^\prime) \delta(t - t^\prime) \label{eq:chi-var-spatial-1}
     \\
    &~&\langle \triangle \chi_W(\mathbf{r}, t)  \triangle \chi_W(\mathbf{r}^\prime, t^\prime)\rangle = D_W \delta(\mathbf{r} - \mathbf{r}^\prime) \delta(t - t^\prime) \;\;\; \label{eq:chi-var-spatial-2}
\end{eqnarray}
with $D_\f, D_W$ being the noise strengths.  For the case of local and instantaneous feedback given by  Eq.~\eqref{eq:kernel-local}, the stochastic version of Eqs.~(\ref{eq:kernel-growth-1} - \ref{eq:kernel-growth-2}) can then be written as: 
\begin{eqnarray}
	 &\partial_t \triangle^2 \phi(\mathbf{r}, t) =  -\alpha \triangle^2 \phi(\mathbf{r}, t) +  \triangle^2\chi_{\phi}(\mathbf{r}, t), & \label{eq:stochastic-growth-1}
	\\
	&\partial_t \triangle^2 W(\mathbf{r}, t) =  -\alpha \triangle^2 W(\mathbf{r}, t) +  \triangle^2 \chi_{W}(\mathbf{r}, t), & \label{eq:stochastic-growth-1}
\end{eqnarray} 
where we substituted Eqs.~(\ref{eq:plate-eqs1}-\ref{eq:plate-eqs2}) to rewrite the left hand side and assumed, for simplicity, that $\alpha_{ij} = \alpha \delta_{ij}$ with $\delta_{ij}$ being the Kronecker delta (see SI for the general case). 

To understand deviations from planarity of the growing lamina, it is natural to consider the fluctuations of the angle between the surface normal ($\hat{N}(\mathbf{r}, t)$) and the vertical ($\hat{x}_3$), which is given by $|\theta(\mathbf{r}, t)| \equiv |\cos^{-1}(\hat{N}(\mathbf{r}, t)\cdot \hat{x}_3)| \approx |\nabla W(\mathbf{r}, t)|$.
In Fourier space, the linear Eqs.~(\ref{eq:chi-var-spatial-2} \& \ref{eq:stochastic-growth-1}) lead to 
\begin{eqnarray}
     & i \omega \hat{W}(\textbf{q}, \omega) = - \alpha \hat{W}(\textbf{q}, \omega) + \hat{\chi}_{W}(\textbf{q}, \omega),  & \label{eq:stochastic-local-W-fourier}\\
        &\langle \hat{\chi}_W (\mathbf{q}, \omega)  \hat{\chi}_W^{\dagger}(\mathbf{q}^\prime, \omega^\prime)\rangle = \frac{(2\pi)^3D_W}{\mathbf{q}^4} \delta(\mathbf{q} - \mathbf{q}^\prime) \delta(\omega - \omega^\prime),&\;\;\;\;\;\; \label{eq:chi-variance-fourier}
\end{eqnarray}
where $\hat{\chi}_W^{\dagger}$ denotes complex conjugation and the Fourier transform of a function $F(\mathbf{r}, t)$ is given by
$\hat{F}(\textbf{q}, \omega) = \int {d^2 \textbf{r} \;dt} \;e^{- i \textbf{q}\cdot \textbf{r} - i \omega t} F(\mathbf{r}, t)$.
This allows us to calculate the strength of the angle fluctuations as 
\begin{eqnarray}
   &~& \langle \theta(\textbf{r}, t)^2\rangle \approx \langle \nabla W(\textbf{r}, t)^2\rangle  = \nonumber \\
   &~& \int \frac{d^2 \mathbf{q}^\prime d\omega^\prime}{(2 \pi)^{3}} \frac{d^2 \mathbf{q} d\omega}{(2 \pi)^{3}} \left(\mathbf{q}\cdot \mathbf{q}^\prime\right) \langle  \hat{W}(\mathbf{q}, \omega) \hat{W}^{\dagger}(\mathbf{q}^\prime, \omega^\prime)\rangle. \;\;\;\; \label{eq:angle-variance-pre}
\end{eqnarray}
Solving Eq.~\eqref{eq:stochastic-local-W-fourier} for $W(\mathbf{q}, \omega)$, substituting it into Eq.~\eqref{eq:angle-variance-pre} and using Eq.~\eqref{eq:chi-variance-fourier}, we get
\begin{eqnarray}
   \langle \theta(\textbf{r}, t)^2\rangle = \int_{L^{-1}}^{h^{-1}} \frac{d^2 \mathbf{q} d\omega}{(2 \pi)^3} \mathcal{P}_{\theta}(\textbf{q}, \omega), \label{eq:angle-fluc-spectral}
\end{eqnarray}
where $\mathcal{P}_{\theta}(\textbf{q}, \omega)$ is the power spectral density given by
\begin{eqnarray}
\mathcal{P}_{\theta}(\textbf{q}, \omega) =  \frac{D_{W}}{\mathbf{q}^2 \left(\alpha^2 + \omega^2\right)}, \label{eq:theta-spectral}\;\;\;\;
\end{eqnarray}
and we assume fluctuations are cutoff for wavelengths smaller than thickness $h$ and larger than system size $L$.  We see that the integral Eq.~\eqref{eq:angle-fluc-spectral} yields $\langle \theta^2(\textbf{r}, t) \rangle \sim \log (L/h)$ and diverges logarithmically, i.e. the ordered flat state is unstable to growth fluctuations for large aspect-ratio laminae. This result does not change for anisotropic feedback, i.e. when we relax the diagonal assumption on $\alpha_{ij}$ (see SI).
\indent \textit{Nonlocal and delayed feedback.---}
Therefore, we ask whether alternative modes of feedback, e.g. those that allow for non-local coupling in space and time, can alleviate the problem of stabilizing the flat state (at linear order). 

%
    	%
%
Perturbations from the flat reference state cause cells to produce signaling molecules (hormones such as auxin) with a delay time scale assumed to be $\Gamma^{-1}$; these propagate diffusively into the local neighborhood \cite{heisler2010alignment, mitchison2015auxin}. Then, a natural model for signal propagation associated with feedback is given by the diffusion equation, whose Green's function satisfies
 \begin{eqnarray}
 \partial_t G_D(\boldsymbol{\rho}, \tau) - \mathcal{D} \triangle G_D(\boldsymbol{\rho}, \tau) = \mathcal{D} \delta(\boldsymbol{\rho}) \delta(\tau), \label{eq:diffusion-greens}
 \end{eqnarray}
 where $\mathcal{D}$ is a diffusion constant (the effect of signal degradation is considered in the SI). If the time scale associated with diffusion ($L^2/\mathcal{D}$), is much smaller than growth time scales ($g^{-1}$), $g L^2 / \mathcal{D} \ll 1$, then the signal concentration approaches equilibrium before the shape changes considerably due to growth. Using the spreading rate of the hormone auxin, we estimate $\mathcal{D} \approx 10 \text{mm}^2/$hour \cite{mitchison2015auxin}, which, along with $L \sim 1$cm and $g \sim 0.01/$hour \cite{sharon-2020}, gives $g L^2 / \mathcal{D} \sim 0.1$. Therefore, we can set the time derivative in Eq.~\eqref{eq:diffusion-greens} to zero, and define the modified Green's function which satisfies $\triangle G_\triangle(\mathbf{r} - \mathbf{r}^\prime) = -\delta(\mathbf{r} - \mathbf{r}^\prime)$. This leads to an extension of the local feedback law in Eq.~\eqref{eq:kernel-nonlocal1} to 
  \begin{eqnarray}
	G_{ij}(\boldsymbol{\rho}, \tau) = -\alpha_{ij} \;\delta(\tau)\; \delta(\boldsymbol{\rho}) - \frac{g e^{-\Gamma \tau}}{h^2} \;\beta_{ij} \;G_\triangle(\boldsymbol{\rho}),\;\;\;\;\;\; \label{eq:kernel-nonlocal1}
\end{eqnarray}
where $g^{-1}$ is the time scale of growth, $\alpha_{ij}$ is a feedback matrix corresponding to the local contribution, $\beta_{ij}$ gives the non-local contribution, and both have units of time$^{-1}$. 
 
 Choosing $\alpha_{ij} = \alpha \delta_{ij}$, $\beta_{ij} = \beta \delta_{ij}$ for simplicity and plugging Eq.~\eqref{eq:kernel-nonlocal1} into Eqs.~(\ref{eq:kernel-growth-1}-\ref{eq:kernel-growth-2}) we get the modified system accounting for nonlocal, delayed feedback as
 \begin{eqnarray}
	  \partial_t \triangle^2 \phi(\mathbf{r}, t) &=&  \triangle^2 \chi_{\phi}(\mathbf{r}, t) -\alpha \triangle^2 \phi(\mathbf{r}, t) \nonumber\\  
	 &+&\frac{g \beta}{h^2} \int_{-\infty}^{t} e^{-\Gamma (t - t^{\prime})} \triangle{\phi}(\mathbf{r}, t^{\prime}) \;dt^\prime, \label{eq:stochastic-growth-nonlocal-1}
	\\
	\partial_t \triangle^2 W(\mathbf{r}, t) &=&  \triangle^2 \chi_{W}(\mathbf{r}, t) -\alpha \triangle^2 W(\mathbf{r}, t) \nonumber\\  
	 &+&\frac{g \beta}{h^2} \int_{-\infty}^{t} e^{-\Gamma (t - t^{\prime})} \triangle{W}(\mathbf{r}, t^{\prime}) \;dt^\prime,\label{eq:stochastic-growth-nonlocal-2}
\end{eqnarray}
 where we used Eqs.~(\ref{eq:plate-eqs1} - \ref{eq:plate-eqs2}) and integrated by parts to obtain the Laplacian feedback appearing on the right hand sides. Here the stochastic terms $\triangle^2 \chi_{W}(\mathbf{r}, t)$ and $\triangle^2 \chi_{\phi}(\mathbf{r}, t)$ satisfy Eqs.~(\ref{eq:chi-var-spatial-1}-\ref{eq:chi-var-spatial-2}) as in the local feedback case.

\begin{figure*}  [ht!]
\includegraphics[width=0.8\textwidth]{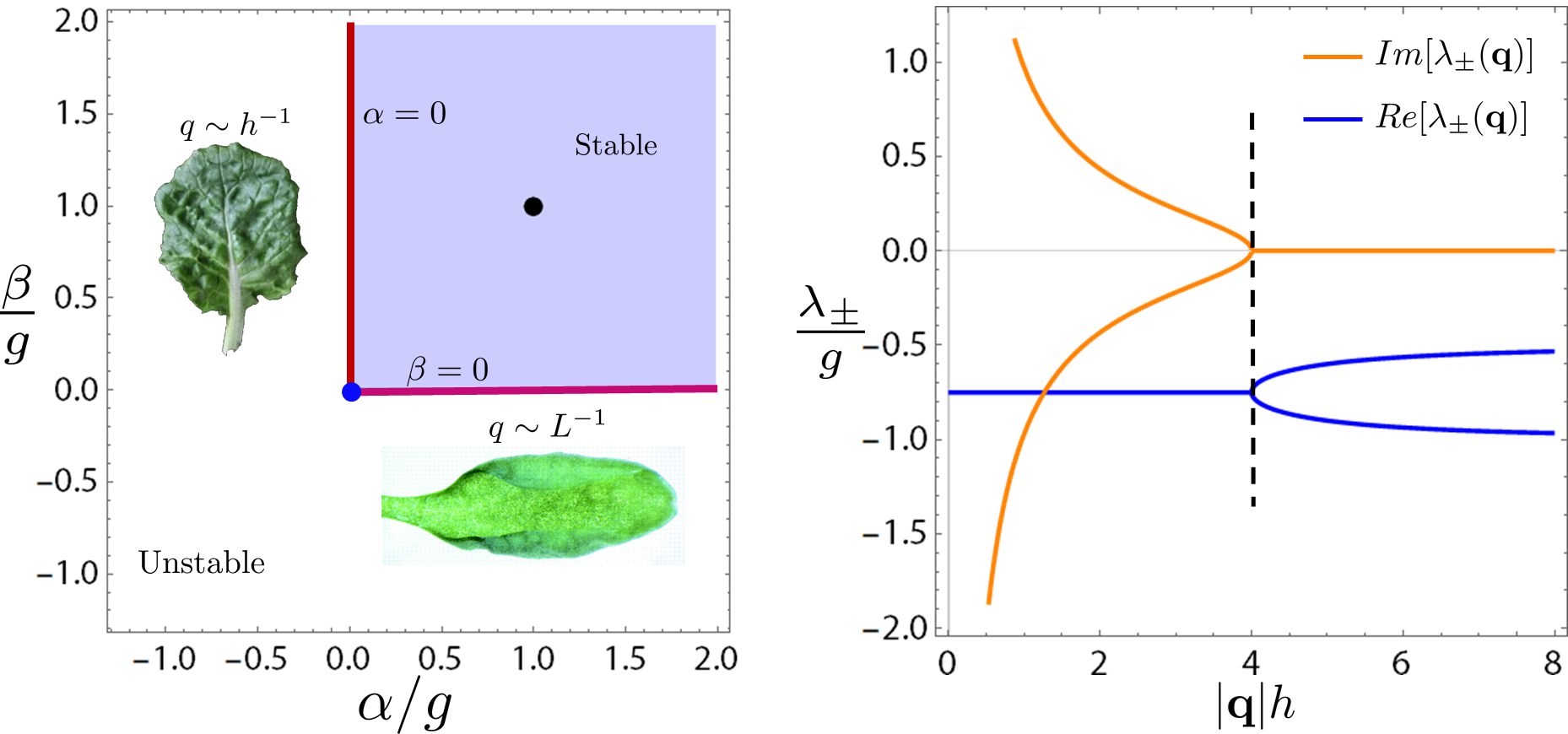}
	\caption{\textbf{Stability phase diagrams:}  (A) The phase diagram corresponding to Eq.~\eqref{eq:combined-rates}. The thick lines represent the boundary of the stable region which are in turn divided according to whether the first unstable modes are large ($|\mathbf{q}| \sim L^{-1}$) or small ($|\mathbf{q}| \sim h^{-1}$) deformations. The blue points where the two boundaries meet corresponds to no feedback $\alpha = \beta = 0$. The two types of unstable modes are exemplified by mutant leaves taken from Refs.~\cite{serrano2000curvata,zhang2020krinkle}. (B) The real and imaginary values of the growth rates, given in Eq.~\eqref{eq:combined-rates}, with parameters corresponding to the black point ($\alpha = \beta = g$) in panel (A) with $\Gamma = 0.5 g$. The dashed line indicates the wavenumber below which the modes are oscillatory. Far to the left of this line ($|\textbf{q}| h| \ll 1$) nonlocal feedback dominates, while far to the right ($|\textbf{q}| h| \gg 1$) local feedback dominates. 
	}  \label{fig:stability-region} 
\end{figure*}

 To understand the stability of the flat state to variations in the growth rates, we use the ansatz $\phi(\mathbf{r}, t) = \phi_0 e^{i \textbf{q}\cdot \textbf{r} + \lambda t}$ and $W(\mathbf{r}, t) = W_0 e^{i \textbf{q}\cdot \textbf{r} + \lambda t}$  and substitute into Eqs.~(\ref{eq:stochastic-growth-nonlocal-1} - \ref{eq:stochastic-growth-nonlocal-2}). We find that the deterministic part of the equation gives
 \begin{eqnarray}
    \lambda = -\alpha - \frac{g \beta}{h^2 \textbf{q}^2} \frac{1}{\Gamma + \lambda}.
 \end{eqnarray}
 When the nonlocal feedback contribution $\beta = 0$ we get $\lambda = -\alpha$, as expected from Eq.~\eqref{eq:local-rates} with $\alpha_{ij} = \alpha \delta_{ij}$. When $\beta \neq 0$, we have two solutions given by
\begin{eqnarray}
	\lambda_{\pm}(\mathbf{q}) = -\frac{1}{2} \left(\alpha + \Gamma \pm \sqrt{(\alpha - \Gamma)^2 - \frac{4 \beta g}{h^2 \mathbf{q}^2}}\right), \label{eq:combined-rates}
\end{eqnarray} 
where the contribution of the nonlocal term $\beta$ dominates for long wavelengths ($\mathbf{q}^2 \ll 4 \beta g / h^{2} \alpha^2$), but is negligible for short wavelengths ($\mathbf{q}^2 \gg 4 \beta g / h^{2} \alpha^2$). We see that the flat configuration is stable only when $\alpha > 0$ (in the limit $\textbf{q}^2 \to \infty$,  $\lambda_{+} = -\alpha$ and $\lambda_{-} = -\Gamma$) and $\beta > 0$ (otherwise when $\textbf{q}^2 \to 0$ one can get a positive growth rate). In Fig.~\ref{fig:stability-region}A, we show the boundary of the stable parameter region indicated by whether the unstable mode has large ($|q| \sim L^{-1}$) or small wavelength ($|q| \sim h^{-1}$), each exemplified by a mutant leaf \cite{serrano2000curvata,zhang2020krinkle}. 
Fig.~\ref{fig:stability-region}B shows the growth rate as a function of wavenumber with parameters chosen in the stable region ($\alpha = \beta = g$). For long wavelengths we generically get oscillations on the way to the flat state (Fig.~\ref{fig:stability-region}B), which may be related to the observed fluttering behavior in leaves during development \cite{avocado-fluttering}.

To understand how fluctuations modify the deterministic feedback dynamics considered above, we start by writing the out-of-plane response of the plate Eq.~\eqref{eq:stochastic-growth-nonlocal-2} in Fourier space as 
\begin{eqnarray}
i \omega \hat{W} = \hat{\chi}_{W}(\textbf{q}, \omega) - \left(\alpha + \frac{g \beta}{h^2 \textbf{q}^2}\frac{1}{\Gamma + i \omega}\right) \hat{W}(\textbf{q}, \omega).&\;\;\;\;\; \label{eq:stochastic-nonlocal-W-fourier}
\end{eqnarray}
Solving Eq.~\eqref{eq:stochastic-nonlocal-W-fourier} for $\hat{W}(\textbf{q}, \omega)$ and using Eq.~\eqref{eq:chi-variance-fourier}, we get an expression for the power spectral density of the normal angle fluctuations given by Eq.~\eqref{eq:angle-fluc-spectral} modified to account for non-local feedback that reads
\begin{eqnarray}
	\mathcal{P}_{\theta} = \frac{(\omega^2 + \Gamma^2) \textbf{q}^2\;  D_{W}}{\left((\omega^2 - \alpha \Gamma) \;\mathbf{q}^2 - \frac{g \; \beta}{h^2}\right)^2 +  \;\mathbf{q}^4 (\alpha + \Gamma)^2  \omega^2}.\;\;\;\; \label{eq:non-local-PW}
\end{eqnarray}
Unlike the case of local feedback (when $\beta=0$) corresponding to Eq.~\eqref{eq:theta-spectral}, when $\beta \ne 0$, the spectral density $\mathcal{P}_{\theta}(\mathbf{q}, \omega)$ in Eq.~\eqref{eq:non-local-PW} is well behaved in the long-wavelength limit and vanishes when $\mathbf{q}^2 = 0$. As a result, angle fluctuations remain finite, $\langle \theta(\mathbf{r}, t)^2\rangle \sim O(L^0)$, as $L \to \infty$. However, in contrast to the case of local feedback, the fluctuations described by Eq.~\eqref{eq:non-local-PW} are not scale invariant ($\mathcal{P}_{\theta}(\mathbf{q}, \omega)$ is not a power law in $\textbf{q}^2$).

To quantitatively compare our model to recent experiments \cite{sharon-2020}, we look at the fluctuations in areal strain rate, $\mathcal{A}(\textbf{r},t) \approx \partial_t \epsilon_{ii}(\textbf{r},t)/2$, where $\epsilon_{ii}$ is the sum of elastic and growth strain tensors, (see SI and \cite{long-leaf}).  The average areal strain rate is defined as $\bar{\mathcal{A}}(\mathbf{r}, t) \equiv \frac{1}{\pi R^2} \int_{R_{\textbf{r}}} \mathcal{A}(\mathbf{r}^\prime, t^\prime) d^2\mathbf{r}^\prime$ and its variance is defined as $\Sigma_\mathcal{A}(R)^2 \equiv  \left \langle \left(\bar{\mathcal{A}}(\mathbf{r}, t) - \langle\bar{\mathcal{A}}(\mathbf{r}, t)\rangle \right)^2 \right\rangle$,  coarse-grained over discs of size $R$ centered at the point  ${\textbf{r}}$ (Fig.~\ref{fig:problem-setup}).

\begin{figure}  [t!]
	\includegraphics[width=\columnwidth]{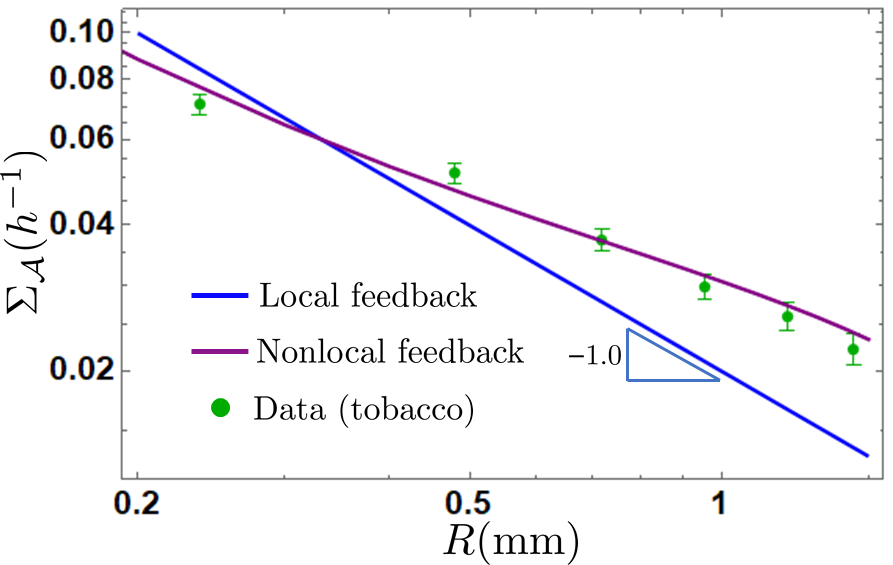}
	\caption{\textbf{Fluctuations in areal growth rate:} 
	 A log-log plot of the standard deviation in areal strain rate $\Sigma_\mathcal{A}$ (in units of inverse hours), where $\mathcal{A}(\textbf{r}, t) \approx \partial_t {\epsilon}_{ii}(\textbf{r}, t)$ is averaged over a region of size $R$ (see Fig.~1). Following Eq.~\eqref{eq:non-local-PA}, local feedback ($\beta = 0$) leads to $\Sigma_\mathcal{A} \propto R^{-1}$  whereas the nonlocal feedback case (with $\alpha = 0, \beta g/h^2 = 1000, T = 0.01, \Gamma = 10$) leads to behavior that can fit the data from Ref.~\cite{sharon-2020} by choosing the noise strength $D_{\phi}$ (see SI for details of parameter fits).
	}  \label{fig:stochastic-growth} 
\end{figure}


To calculate $\Sigma_\mathcal{A}(R)$ using our model, we note that $\mathcal{A}(\textbf{r},t) \sim \partial_t \triangle \phi(\textbf{r},t)$, since $E \triangle\phi(\textbf{r},t)$ is the trace of the elastic stress tensor. As Eq.~\eqref{eq:stochastic-growth-nonlocal-2} has a similar form to Eq.~\eqref{eq:stochastic-growth-nonlocal-1}, we can repeat the steps leading to Eq.~\eqref{eq:non-local-PW} (see SI) to obtain the result
\begin{eqnarray}
   &~& \Sigma_\mathcal{A}^2 \propto \langle \partial_t \triangle \phi(\textbf{r},t)^2\rangle \propto \int^{R^{-1}, T^{-1}} \frac{d^2 \mathbf{q} d\omega}{(2 \pi)^3} \mathcal{P}_{\mathcal{A}}(\textbf{q}, \omega),\;\;\;\;\;\;\;\; \label{eq:non-local-PA-0}\\ 
   &~&  \mathcal{P}_{\mathcal{A}} \propto \frac{\omega^2 (\omega^2 + \Gamma^2) \textbf{q}^4\;  D_{\phi}}{\left((\omega^2 - \alpha \Gamma) \;\mathbf{q}^2 - \frac{g \; \beta}{h^2}\right)^2 +  \;\mathbf{q}^4 (\alpha + \Gamma)^2  \omega^2}, \label{eq:non-local-PA} \;\;\;\;
\end{eqnarray}
where the $\textbf{q}$ integral is cutoff by $R^{-1}$ due to spatial averaging over the disk of size $R$, $T^{-1}$ is the high frequency cutoff (see SI), and the additional factor of $\mathbf{q}^2 \omega^2$ compared with Eq.~\eqref{eq:non-local-PW} is due to the different number of derivatives in $\partial_t \triangle \phi(\textbf{r},t)$ and $|\nabla W(\textbf{r},t)|$. 

For purely local feedback ($\beta = 0$), $\mathcal{P}_{\mathcal{A}}(\mathbf{q}, \omega) = \mathcal{P}_{\mathcal{A}}(\omega)$ is independent of $\textbf{q}$ and we obtain $\Sigma_\mathcal{A} \propto R^{-1}$ (as can be seen by changing coordinates $q \to R q$ in the integral Eq.~\eqref{eq:non-local-PA-0}). 
Experiments in \cite{sharon-2020} give $\Sigma_\mathcal{A} \propto R^{-p}, p \approx 0.61$ (Fig.~\ref{fig:stochastic-growth}), which interestingly differs from the $R^{-1}$ behavior expected if growth fluctuations were not correlated spatially, indicating a nonlocal nature in the feedback law or long-range correlations in the fluctuation spectrum. For purely nonlocal feedback ($\alpha = 0$), we do not get power-law behavior (because the power spectral density is not scale invariant unless $\Gamma=0$, in which case we have to consider the range of the integral in ~\eqref{eq:non-local-PA-0} to ensure non-divergent behavior).   In Fig.~\ref{fig:stochastic-growth} we show the results obtained by integrating Eq.~\eqref{eq:non-local-PA-0} numerically with specific choices for the parameters (see figure caption and SI for details) and see that they better capture the experimental observations from \cite{sharon-2020}. 

\textit{Conclusion.---} 
Growing a flat lamina stably is difficult because small fluctuations in metric and curvature growth are both destabilizing on long length scales. We formalize this intuitive result in terms of a simple mathematical framework that couples elasticity and strain-induced feedback and show that local and instantaneous feedback is insufficient to stabilize long wavelength buckling modes [Eq.~\eqref{eq:theta-spectral}]. In contrast, spatially nonlocal, temporally delayed feedback suppresses these long wavelength fluctuations [Eq.~\eqref{eq:non-local-PW}] and better captures experimentally observed scaling behavior [Fig.~3B]. Natural extensions of this work include generalizing the results to elastic shells and into the nonlinear regime. 

\bibliography{references}

\end{document}